\documentclass[conference]{IEEEtran}
\IEEEoverridecommandlockouts
\usepackage{cite}
\usepackage{amsmath,amssymb,amsfonts}
\usepackage{algorithmic}
\usepackage{graphicx}
\usepackage{textcomp}
\usepackage{xcolor}
\def\BibTeX{{\rm B\kern-.05em{\sc i\kern-.025em b}\kern-.08em
    T\kern-.1667em\lower.7ex\hbox{E}\kern-.125emX}}
\begin{document}

\title{Embodied Cognition Guides Virtual-Real Interaction Design to Help Yicheng Flower Drum Intangible Cultural Heritage Dissemination\\

\thanks{This research was partially supported by the National Social Science Foundation of China (20CJY045) and the Chinese Academy of Management Sciences (KJCX25708).}
}

\author{\IEEEauthorblockN{1\textsuperscript{st} Yuhan Ma\#}\thanks{\#These authors contributed equally to this work.}
\IEEEauthorblockA{\textit{Beijing Forestry University} \\
Beijing, China \\
18500619089@163.com}
\and
\IEEEauthorblockN{2\textsuperscript{nd} Weiran Zhao\#}
\IEEEauthorblockA{\textit{Beijing Film Academy} \\
Beijing, China \\
wr.zhao@outlook.com}
\and
\IEEEauthorblockN{3\textsuperscript{rd} Xiaolin Zhang}
\IEEEauthorblockA{\textit{
University of Auckland} \\
Auckland, New Zealand \\
xiaolinzhang1001@gmail.com}

\and
\IEEEauthorblockN{4\textsuperscript{th} Ze Gao*} \thanks{*Ze Gao is the corresponding author.}
\IEEEauthorblockA{\textit{
Hong Kong University of Science and Technology} \\
Hong Kong SAR, China \\
zegaoap@hotmail.com}
}

\maketitle

\begin{abstract}

In order to make the non-heritage culture of Yicheng Flower Drum more relevant to the trend of the digital era and promote its dissemination and inheritance, the design and application of gesture recognition and virtual reality technologies guided by embodied cognition theory in the process of non-heritage culture dissemination is studied. At the same time, it will enhance the interaction between people and NRM culture, stimulate the audience's interest in understanding NRM and spreading NRM, and create awareness of preserving NRM culture. Using embodied cognition as a theoretical guide, expanding the unidirectional communication mode through human-computer interaction close to natural behavior and cooperating with multisensory information reception channels, so as to construct an embodied and immersive interactive atmosphere for the participants and enable them to naturally form the cognition and understanding of the traditional culture in the process of interaction. The dissemination of the non-heritage culture Yicheng Flower Drum can take the theory of embodied cognition as an entry point, and through the virtual and real scenes of Yicheng Flower Drum and the immersive experience, we can empower the interaction design of non-heritage culture dissemination of the virtual and real, and provide a new method for the research of digital design of non-heritage culture.

\end{abstract}


\section{Introduction}

Intangible cultural heritage (ICH) is a vital aspect of China's traditional culture, with a significant role in global cultural exchange. However, in the digital age, preserving and spreading traditional non-heritage culture faces challenges. Yicheng Flower Drum, a part of China's National Intangible Cultural Heritage, is struggling due to multicultural influences. To adapt to the modern era and seize opportunities for non-heritage cultural development, we need contemporary communication methods. Digital technology has emerged as a primary tool for sharing non-heritage culture, addressing the limitations of traditional protection methods. Scholars have increasingly recognized the value of digital means in promoting non-heritage culture\cite{b1}. Ma Xiaona and other scholars have discussed the specific technical means and current operation form of digitization of non-heritage culture, and have made a summary and outlook of research needs and development trends; Wei Juntao and others have discussed the specific technical means and current operation form of digitization of non-heritage culture\cite{b2}. Taking the non-heritage Ten Mile Red Makeup as an example, based on the digital twin theory using tilt photogrammetry, the transformation of the digital design of the non-heritage was practically verified; Zhang Nana et al. \cite{b3} proposed the concept of a multidimensional field to explore the digital means of NRH display. Current research often prioritizes technical innovation and visual presentation, overlooking the need for a strong theoretical foundation and context-specific communication technology for non-heritage culture. This paper starts with embodied cognition theory and integrates emerging technologies such as virtual reality and gesture recognition. It enhances digital non-heritage culture communication in perception, behavior, material, and emotion dimensions, aiming to protect Yicheng Flower Drums. This study offers insights into non-heritage culture communication methods and related research.

\section{Explanation of relevant concepts}

\subsection{Intangible Cultural Heritage}

Intangible cultural heritage, born from centuries of social life and production, has profound historical, cultural, scientific, and artistic value. Safeguarding and passing down these cultural items is crucial to preserving, evolving, and sharing China's rich traditional culture and fostering a stronger national identity\cite{b4,b16}. As a living cultural heritage oriented to people, it is based on practices such as oral transmission and teaching by example. In today's evolving society, non-heritage culture requires innovative approaches for preservation, adaptation, and wider dissemination. Modern design can serve as a medium to enhance protection, inheritance, and development, ensuring the continuity and relevance of this centuries-old tradition.

\subsection{Yicheng Flower Drums}

The art of drum dancing, which is a combination of ancient and modern, has shined brilliantly in the long history of the Chinese nation and is an important part of Chinese folk culture and art. Yicheng flower drum, as a member of the very special characteristics of the form, is colorful, lively, and of bright rhythm and magnificent\cite{b5}. With a distinctive performance style that blends singing, dancing, warm emotions, exaggerated movements, and simplicity, Yicheng in Shanxi Province is renowned as the 'Hometown of Flower Drums.' These drums have a rich cultural heritage, earning them a place in the first wave of national non-heritage cultural protection projects in 2006.

With digitization, Yicheng flower drums face succession challenges and dissemination limitations tied to local customs. Preserving and spreading them is crucial for local life, culture, and festivals. To innovate while preserving their essence, Yicheng Flower Drums should adapt to the digital age, expand their audience, and achieve creative transformation.
\subsection{Possessive Cognition}
Embodied cognition theory explores the intricate connections and interactions between cognitive processes and the perceptual and motor systems. Traditionally, human thought was often regarded as an abstract concept with limited connection to the physical environment. However, with the continuous advancement of artificial intelligence technology, researchers have gained fresh insights into embodied cognition through the lens of AI's operational logic. This paradigm shift suggests a strong interplay between cognitive processes and the external physical environment, highlighting that the perceptual and motor system serves not only as a vessel for interaction but also as an integral component of cognition itself. This article synthesizes the core tenets of embodied cognition, as elucidated by Margaret Wilson, emphasizing that cognition is situated, temporally constrained, reliant on environmental resources, intimately connected with the physical surroundings, oriented towards action, and intrinsically rooted in bodily experiences\cite{b6}.

\begin{figure}[h]
  \centering
  \includegraphics[width=\linewidth]{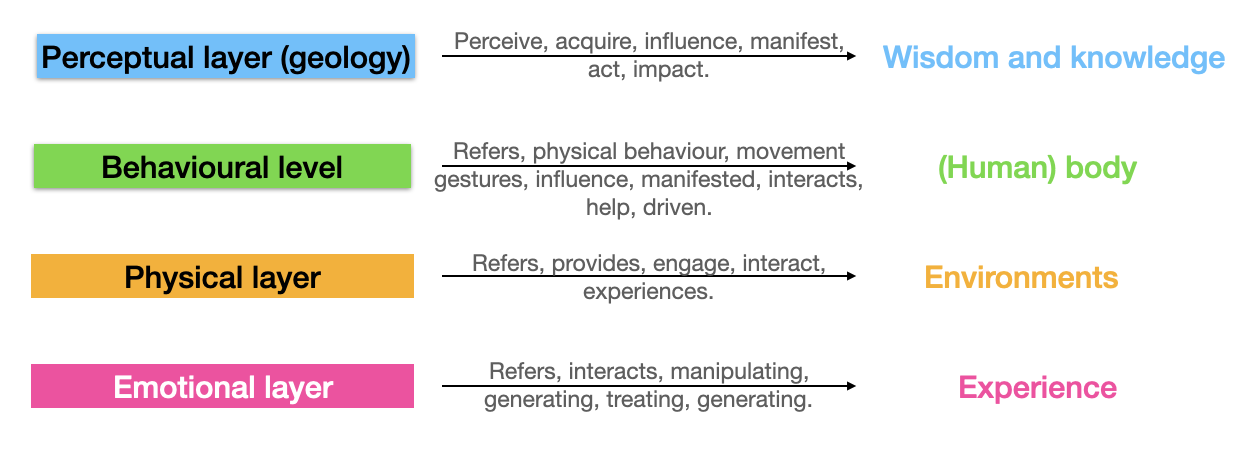}
  \caption{Hierarchy of embodied interactions affected by digital means technology.}
  \label{fig1}
\end{figure}

In 'The Phenomenology of Perception,' Merleau-Ponty argues that the human body is inseparable from the world, just as the heart is within the body. This idea forms the foundation of Embodied Cognition, which emphasizes the interaction between the brain, body, and environment in human cognition. It underscores the importance of bodily participation in cognitive activities and the vital role of the surrounding environment and bodily actions in cognitive and emotional well-being. Based on the theory of embodied cognition, the author proposes that the embodied interaction under the influence of virtual reality and other digital technology means are, respectively, expressed as the perception layer "mind", behavior layer "body", material layer "environment", emotional layer "environment", "physical layer", "physical layer", "physical layer", "physical layer", "physical layer", "physical layer", "physical layer", "physical layer", "physical layer", and "emotional layer". The four factors of "mind" in the perceptual layer, "body" in the behavioral layer, "environment" in the material layer, and "experience" in the emotional layer are proposed (see Figure \ref{fig1}).

With the technical support of artificial intelligence and human-computer interaction, the process of non-heritage culture dissemination based on the theory of embodied cognition can enable the audience to change the traditional passive state of receiving information and enable users to obtain the real experience of being in the virtual scene through movements, gestures, and other interactive behaviors to enhance the degree of participation and creativity. At the same time, the design under the guidance of embodied cognition closely matches the cognitive process and logic of the users, and truly integrates the mind, body, and environment, so as to obtain the transmission and dissemination of the knowledge of non-heritage culture, and ultimately to achieve the sublimation and enhancement of the users' emotions.

\subsection{Gesture Recognition}

Gesture recognition is a human-computer interaction method based on capturing and interpreting hand movements for communication with computers. It offers a more intuitive and user-friendly experience than traditional keyboard and mouse input, with faster learning curves. Users can engage their body movements, achieving embodied cognition when combined with virtual reality technology. Today, most gesture recognition relies on vision-based solutions, which are cost-effective and less restrictive than traditional data gloves. Gesture recognition can be categorized into two-dimensional gesture recognition, two-dimensional hand recognition, and three-dimensional gesture recognition.

Two-dimensional gesture recognition interprets user gestures in the XY plane using 2D image input, primarily for simple command recognition. In contrast, 3D gesture recognition adds the Z-axis, capturing depth information for gestures in 3D space. This is more suitable for virtual spaces and complex interactions. Two-dimensional recognition excels in tracking both static and dynamic gestures, making it ideal for complex hand movements and providing richer human-computer interactions, like drawing on-screen or responding quickly to prompts. Gesture recognition typically involves four key steps: gesture acquisition (capturing user gestures), gesture segmentation (separating foreground and background), gesture analysis (estimating parameters), and gesture recognition (matching and extraction) \cite{b7}. The last step is to match and extract the gestures through feature value extraction.

The application of gesture recognition technology in the interactive exhibition design of the Yicheng Flower Drum can play a better role in assisting the learning and inheritance of non-heritage culture. As procedural knowledge, non-heritage culture mainly relies on subconscious experience and systematic movements as the carrier, which is difficult to express and taught by words. According to the theory of embodied cognition, the structure of the body and sensory movement play an important role in the formation of cognition, and the origin of cognitive structure is the general coordination of movement, which is the initial contact surface and contact mode of the body and the environment in the interactive activities, and serves as a bridge between the subject and the object\cite{b8}. Gesture recognition technology enhances our understanding of the relationship between the body and the environment, allowing the audience to engage with Yicheng Flower Drum through movement and gesture interactions. Participants project their perceptions onto virtual bodies, responding to program prompts with corresponding dance movements. This fosters embodied cognition, enabling participants to learn Yicheng Flower Drum movements in a game-like process.

\subsection{Virtual Reality}

Virtual Reality, as the earliest and more far-reaching basic classification under the Extended Reality Technology, is a solution for constructing incarnate immersive environments with Imagination, Interaction, and Immersion as the main technical characteristics. It refers to computer systems used to create artificial worlds in which users feel immersed, and can roam around and manipulate objects\cite{b9}. "Virtual reality offers a heightened immersive experience, engaging the audience on multiple sensory levels. It achieves immersion by expanding visual coverage through head-mounted displays or similar devices, blurring the line between the audience and the screen. Traditional media rely on imagination and conscious projection into the narrative, while virtual reality's immersion is enhanced by its natural three-dimensional human-computer interaction. VR mimics real-life interactions, allowing users to engage using HMDs, gloves, or controllers, creating a strong sense of presence. It fosters embodied cognition, where users adapt to the virtual environment and project their perception onto virtual bodies, driving active cognitive engagement and enhancing information reception."

Dongyan Nan, in their bibliometric analysis of the academic framework of the metaverse, identified that terms such as "Virtual reality," "Virtual worlds," "Augmented reality," "Extended reality," and "Mixed reality" exhibit a high frequency of occurrence. This observation signifies a strong association between these terms and the concept of the metaverse. Furthermore, it suggests that the metaverse is a comprehensive and multifaceted concept within academic discourse\cite{b10}.

In summary, in the field of interactive exhibition, the application of virtual reality technology can provide participants with more realistic scenes and real-time feedback of the interactive process and put them in a real situation. Mobilizing the enthusiasm of the participants, helps them to realize the effective transfer of knowledge\cite{b11}. Yicheng Flower Drums Because of its intangible cultural heritage, Yicheng Flower Drum has greater limitations in the inheritance process, and through the creative transformation and innovative development of virtual reality technology in the form of dissemination, the participants can realize the subtle appreciation of Yicheng Flower Drum's performance method and cultural value in the virtual embodied environment.

\section{Design Principles of Yicheng Flower Drum Immersive Interactive Experience under the Theory of Embodied Cognition}

Relying on the development of science and technology and innovative digital technology means, that the human modern lifestyle brings great convenience at the same time, but also to the inheritance and development of non-heritage culture to bring new opportunities. Yicheng Flower Drums have been passed down through thousands of years of history, and have not only evolved into a unique folk dance, distinctive and extraordinary. With the aging of performers and audiences and the loss of young audiences, the Yicheng Flower Drum is in urgent need of new means to innovate communication methods and improve communication efficiency. Modern design is paying more and more attention to user-centered user experience, which is the key issue to be solved in the communication of Yicheng Flower Drum.

\subsection{Energising the User's Cognitive State}

\begin{figure}[h]
  \centering
  \includegraphics[width=\linewidth]{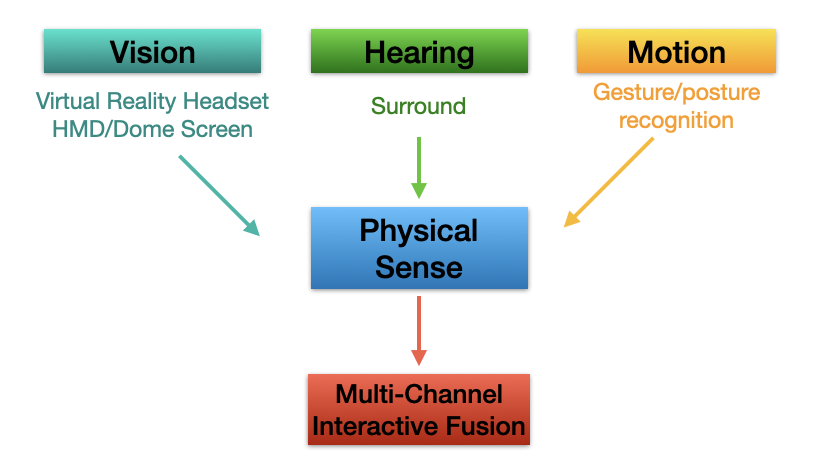}
  \caption{Yicheng flower drum immersive interactive work design flow chart.}
  \label{fig2}
\end{figure}

The theory of embodied cognition advocates the integration of body and mind and attaches importance to the effect of the body in the cognitive process of human beings\cite{b12}. In this process, the organs of the body perceive external things and then transmit the resulting information to the brain, forming instinctive embodied cognition. This means that the communication method should be close to the physical characteristics and structure of the human body, and close to the user's natural interaction mode. It effectively breaks the interaction constraints brought by a single interaction mode, and the interactions of multiple channels are integrated together to enable the audience to achieve the most natural and efficient two-way human-computer interaction, see Figure \ref{fig2}.

In this study, users engage in interactive experiences with the assistance of immersive interactive devices, contributing to the improvement of user experience and further enhancing the overall sense of immersion. Drawing from Seungjong Sun et al.'s research on player motivation in Massively Multiplayer Online Role-Playing Games (MMORPGs), it is posited that immersion motivation and achievement motivation play pivotal roles in enhancing the overall enjoyment of the gaming experience. Conquering challenging tasks within the game is shown to heighten the player's satisfaction and enjoyment. The immersive nature of role-playing games, such as MMORPGs, can trigger the Proteus effect, thereby amplifying the sense of immersion and motivation to overcome challenges. In terms of interaction design, this study aligns with established principles from traditional game design. Participants actively control virtual characters, guiding them through various action-based challenges during the interactive process\cite{b13}.

Traditional Yicheng Flower Drum communication tends to treat the audience as a purely visual receiver, and there is a lack of practice and examination of whether the audience has truly internalized the performance content, aroused cognitive interest, and formed deep memories\cite{b14}. The Yicheng Flower Drum Communication Method Different from traditional performance forms, audiences can now rely on various digital means to apply embodied cognition theories to communicate and interact with each other, forming a natural intuitive way to obtain a good multi-sensory experience, thus stimulating cognitive behaviors and states. For the target audience, what they see are mostly singing and dancing actions and a lively atmosphere. Due to the conditions of time, geography, and space, the audience cannot experience and spread them in person. Integrating the theory of embodied cognition into the communication process of Yicheng flower drums and other non-heritage cultures can help the audience to be brought into the relevant situation, and turn the one-way and unique communication mode into a multi-way and multi-channel form of practical interaction. Through the experience, they can gain embodied cognitive experience, increase cognitive interest, and form deeper memories.

\subsection{Behavioural Imitation Promotes Cognitive Level}

The theory of embodied cognition emphasizes that the input and output links of information are connected by bodily movements and that bodily manipulation of movement is also an important vehicle for cognitive behavior. That is, cognition is mimetic, and when an audience perceives an object and performs behavioral mimicry, the process can facilitate motor, social, and cognitive effects. In the cognitive process, during the brain's processing of information formed by the body and the environment in which the body is located, the limbs produce corresponding behavioral activities, which are episodic behaviors produced by the brain's integration of information\cite{b15}. This is the outward behavior of the brain after integrating the information. After the information is perceived, the body produces interactive behavioral activity experience, the behavioral experience forms memory cognition in the brain, and the memory cognition in the environment will make the body produce behavioral activities, and finally form the behavioral embodied cognition in the environment. As a result, the audience's gestures, movements, and techniques are very important in acquiring information and outputting actions. Similarly, in order to transform the traditional viewing mode of Yicheng Flower Drums, the important principle is to make the audience actually experience and follow the guidance to imitate the movements, and gradually construct the cognition of Yicheng Flower Drums of the non-heritage culture from the physical experience.
Based on the theory of embodied cognition, the Yicheng Flower Drum Experience Interactive Design integrates the performance scene, performance movements, and environmental atmosphere using somatosensory interactive technology, and upgrades the flower drum movements to a high level of cultural dissemination through the audience's imitation, which puts into practice the concepts of "user-centredness" and "edutainment for fun". The concepts of "user-centered" and "edutainment" are put into practice. In the intelligent computing era and with the support of human-computer interaction technology, the interactive communication of Yicheng Flower Drum based on embodied cognition enables users to present information through audio-visual media in the real space and to experience the real atmosphere and performance forms in the virtual space with the help of interactive devices.

\vspace{-0.5mm}
\subsection{Virtual atmosphere creates cognitive emotions}

In the cognitive process, the information produced by people's bodies in the environment stimulates the brain and then realizes embodied cognition. In this process, in addition to completing the input of information, the audience also carries out emotional activities for the corresponding information, such as national pride and identity. The communication form of Yicheng Flower Drum, a non-heritage culture, should also pay attention to the emotional experience caused by local and overall scenarios shaping the audience's cognition. On the one hand, designers can add interesting and relevant elements at the visual level to stimulate the audience's curiosity and gain emotional resonance, while paying attention to avoiding overly complex interaction methods, which may make the audience feel bored. On the other hand, the interaction design should be combined with the audience's experience and the humanistic environment, so that the audience can emotionally generate a sense of empathy and a sense of belonging to this non-heritage culture, and obtain a high level of emotional experience.

\section{Gesture Recognition Based Immersive Interactive Work Design - The Case of Yicheng Flower Drums}

\subsection{Design Process}

This study mainly takes Yicheng Flower Drum, an intangible cultural heritage of Shanxi Province in China, as a research case to explore the specific application of immersive and gesture recognition technology in the process of protection and inheritance of intangible cultural heritage in China. Yicheng Flower Drum is unique in its musical instruments, costumes and performance forms, and its history has a long history, which is recorded in the Yicheng County Records in the "Volume of State Relatives", "Volume of Sacrifices", "Volume of Rituals and Customs" and so on. In the "Yicheng County Zhi", there are records of flower drums in the "Guoqi Volume", "Ritual Volume", "Ritual Volume" and so on.

Under the background of the contemporary rapid development of the Internet, the fragmentation and de-ideologization of the communication media are gradually dissolving the mode and form of communication of intangible cultural heritage. Combining the communication and interaction methods that mainstream groups are accustomed to accepting is particularly important for the protection and dissemination of intangible cultural heritage, therefore, this study will adopt gesture recognition combined with a virtual reality helmet (HMD) and immersive dome screen for the innovative presentation of Yicheng Flower Drums in the design of the specific program. Due to the differences between different audience groups, the immersive dome screen and HMD in this work will provide participants with a choice of visual presentation methods.

\begin{figure}[h]
  \centering
  \includegraphics[width=\linewidth]{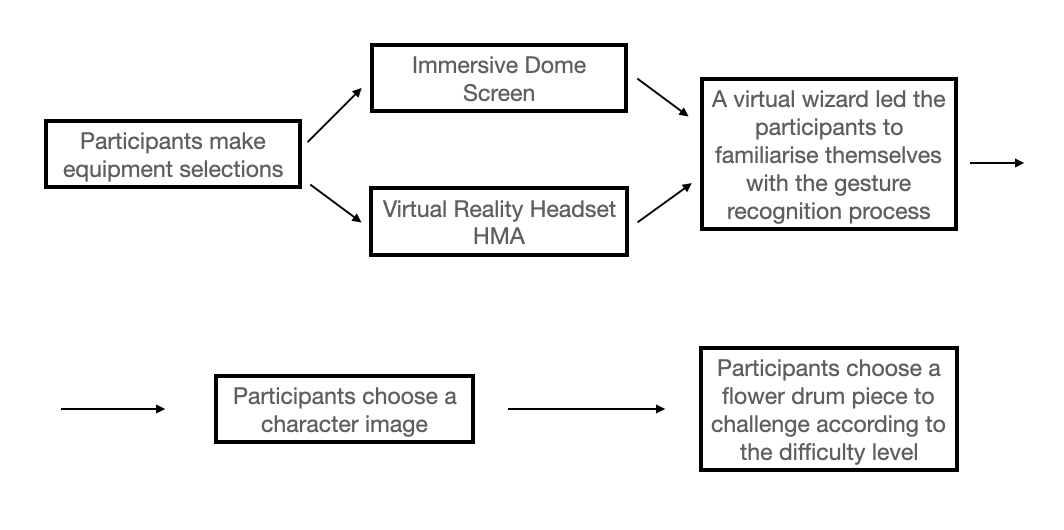}
  \caption{Yicheng flower drum immersive interactive work design flow chart.}
  \label{fig3}
\end{figure}

The main interactive process of this interactive work is as follows: the user first enters the immersive dome screen to participate in the interaction or enters the immersive virtual space by wearing an HMD. When the user enters the virtual space according to the chosen method, the interactive program will provide the participant with a virtual environment built according to the local characteristics of Yicheng, and the digital human guide in the virtual environment will help the audience to gradually familiarise themselves with the interaction process of gesture recognition through a series of guiding processes. The participant chooses the character he/she wants to use by gesture interaction. Subsequently, the participant can choose the name of the corresponding flower drum work to challenge, through the gesture recognition equipment for collection, and follow the prompts to make the corresponding dance gestures, when the gesture and timing of the action and the dance rhythm of the matching, the interactive program will give the corresponding score reward, and in the virtual environment of the surrounding characters will be based on the performance of the participant to make the behavioral feedback such as cheering, applauding, and so on (see Figure \ref{fig3}). The participants can understand and master the movement of the Yicheng Flower Drum in a gradual and orderly way.

In this research, UnrealEngine is mainly used in the interaction program, which is developed together with the gesture recognition device Kinect and the virtual reality device HTC VIVE. The gesture recognition technology is mainly accomplished with the help of the motion capture device Kinect, which is an interactive somatosensory device introduced by Microsoft, mainly composed of an infrared emitter, infrared camera, and color RGB camera. During the gesture recognition process, the infrared emitter emits infrared light to cover the capture area, and then the RGB camera captures the infrared-covered area, and the surface of the infrared-covered area will form randomly distributed diffraction spots. The depth of the image is calculated by the Kinect's built-in processor to obtain information about the position of the tracked object in three-dimensional space, which is used to determine the participant's movement and gesture presentation.

\subsection{Scene Construction}

The theme of this research is centered around the Yicheng Flower Drums, in order to facilitate the interactive experience of the participants, and therefore in the construction of the scene to take a broader field as the main, with a small number of Shanxi traditional houses around the construction to constitute an immersive environment, the construction of traditional houses is mainly based on the relevant picture data for the restoration. In addition to the building, the virtual scene will also be distributed by the virtual digital people composed of residents and the audience with the participants of the interactive feedback, at the same time, the project in the scene lighting will simulate the reality of the natural light lighting, in the process of the participants of the interactive process, the scene of light and shadow with the interactive length of time to produce the corresponding changes.

The virtual characters in this study will be created using the Metahuman platform provided by Epic Games. The Metahuman Creator solution greatly reduces the technology and cost of digital human creation. Compared to the traditional next-generation process of creating virtual characters, Metahuman Creator can complete the construction of high-precision virtual characters, as well as the process of binding bones and controllers to the virtual character's body and face in a short period of time and at a faster speed. Thanks to Metahuman, the project was able to quickly build the digital assets of the virtual audience characters and complete the skeleton assembly in a short period of time through facial scanning or re-creation and blending based on character presets.

\subsection{Interaction Design}

\begin{figure}[h]
  \centering
  \includegraphics[width=\linewidth]{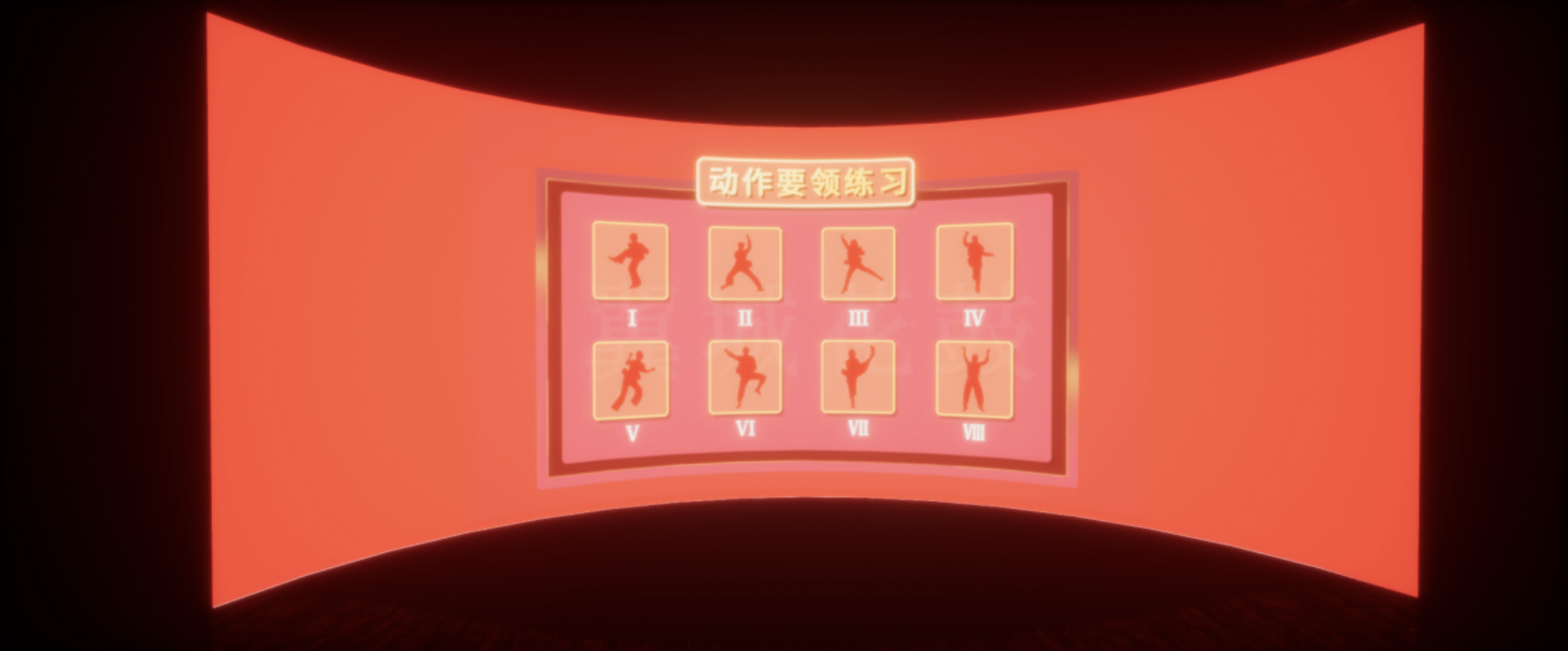}
  \caption{The scene shows the key points of Yi cheng flower drum movement.}
  \label{fig4}
\end{figure}

As an essential component of human-computer interaction, the interaction interface plays a significant role. A good interactive interface can effectively convey information while minimizing the user's learning cycle and learning cost, therefore, this study follows the principle of metaphorical interaction design in the design of the interactive interface, i.e., using graphics as the main constituent element of the interactive interface, presenting icons in an anthropomorphic form to enable participants to quickly master the use of the interactive interface and simplifying the interactive interface as much as possible in terms of presentation. In terms of presentation, the interactive interface is simplified as much as possible, and the layout is designed according to the primary and secondary relationships of menus and functions (see Figure \ref{fig4}).

\begin{figure}[h]
  \centering
  \includegraphics[width=\linewidth]{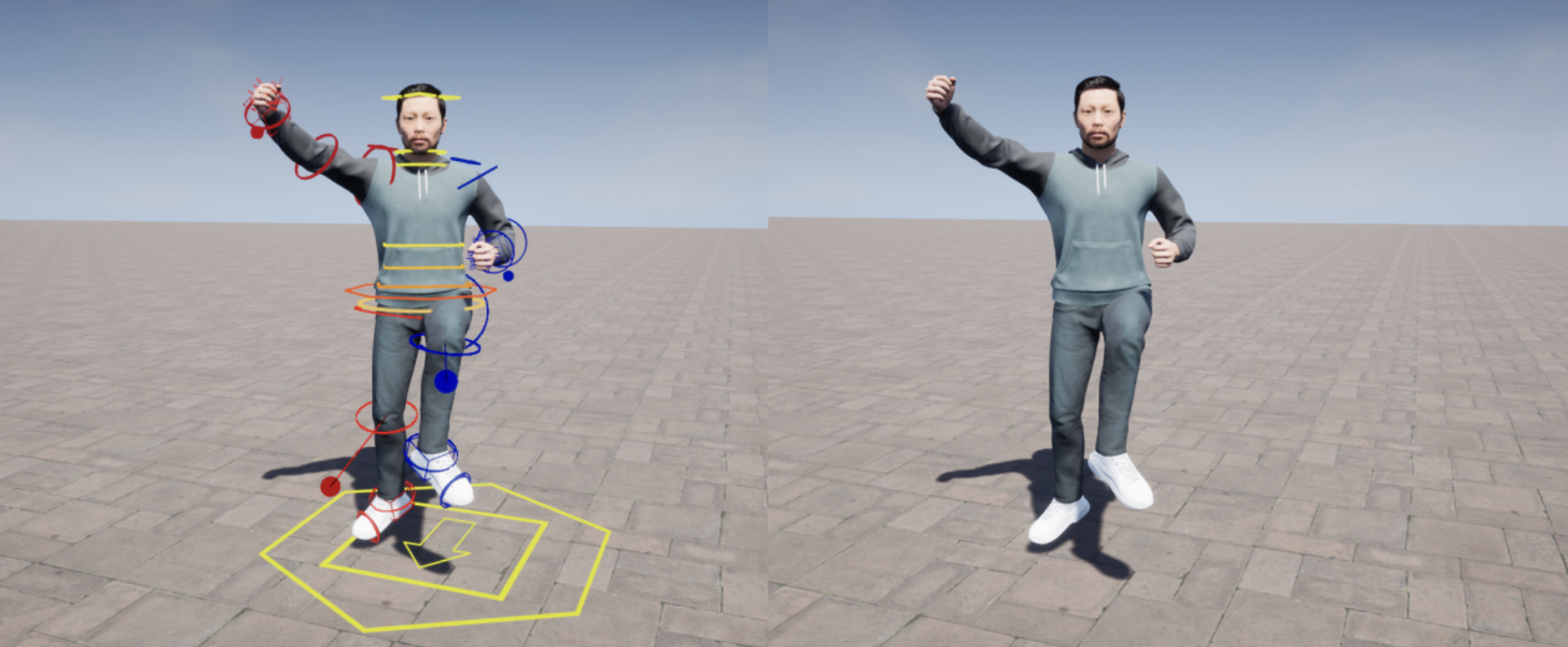}
  \caption{Animation Production and Collection.}
  \label{fig5}
\end{figure}

In order to shorten the animation production cycle and make the performance of the virtual character more vivid and natural during the interaction process, this study will focus on inertial motion capture in animation production and complete the animation production with the manual keyframe adjustment in the later stage of the program. The motion capture process mainly adopts Xsens with FaceWareStudio to perform motion capture in the Unreal Engine and synchronizes the Xsens inertial motion capture with the help of MVN in real time to the Unreal Engine for recording by means of LiveLink (see Figure \ref{fig5}). The study invites the Yicheng flower drummers to perform segmented movements of the dance in motion capture costumes equipped with Xsens inertial motion capture sensors, while the team members perform and record the dance using the inertial motion capture sensors. At the same time, team members used a similar method to provide motion capture data capture for the virtual audience and residents. After the animation data was captured, the Unreal engine was exported to the 3D software for animation restoration to make the characters move more smoothly and naturally.

The position of the bone points in the participant's skeleton and gesture movements were obtained through the motion data captured by Kinect. Yicheng Flower Drums pay more attention to the performance of leg movements, so this study relies on the body sensing equipment to obtain the relative positions of the bone joints of the participants' bodies for gesture recognition and calculates the accuracy of the participants' movements by comparing them with the standard movements of the dancers. Based on this, the scores of the participants' postures and movements are evaluated, and when the postures or movements presented by the participants are more in line with the captured standard movements, the animation state of the virtual audience surrounding the participants will be switched from standby to cheering or applauding in order to enhance the participants' embodied immersion, and the participants will also observe the corresponding score changes through the HMD or through the spherical screen. When the user's score in the challenge reaches a certain level, the program will unlock the next dance essentials exercise for the participant. In order to unlock the next challenge, the audience needs to follow the guidance in the interactive process and continue to imitate the standard gestures; in the process of imitation, the body will gradually build up an embodied cognition from its experience so that the understanding of the Yicheng Flower Drums can be transformed from a viewing to a physical understanding.

For sound production, our project uses sound mimicry using props and common objects to simulate applause, cheers, and environmental noises. We have recorded and categorized various sound clips. During the interaction, when a specific sound condition is triggered, the program selects and plays corresponding sound clips randomly from the library to prevent auditory fatigue. We also utilize Unreal Engine's sound system to simulate sound attenuation and spatialization, enhancing the immersive ambiance.

\section{Conclusion}

Since the 21st century, the state's attention to intangible cultural heritage has been gradually strengthened, and compared with tangible cultural heritage, intangible cultural heritage is more difficult to protect and pass on. As the crystallization of human wisdom and creativity, it carries behind it the inheritance of national character and spirit and highlights the wisdom and life attitude of the predecessors, thus, it has an unignorable role in maintaining the national characteristics and carrying forward the spirit of Chinese culture. The inheritance of intangible cultural heritage requires not only the recording of the media but also the personal understanding of every citizen of the Chinese nation. This study takes Yicheng Flower Drum, a representative intangible cultural heritage of Shanxi, as the research object, combines the theory of embodied cognition with the creative transformation of the communication method, and expands the unidirectional communication method into an immersive interactive form of embodied immersion by combining virtual reality and gesture recognition, which is easier to accept by the participants and at the same time makes the participants mobilize their bodies in a way that conforms to their natural behaviors to achieve the goal of mind, body, and spirit. At the same time, it is easier for the participants to accept and also enables them to mobilize their bodies in a way that is in line with natural behaviors, so as to achieve the integration of mind, body, and environment, and to improve the audience's participation, thus enabling participants of different ages to gradually appreciate the charms of the Yicheng Flower Drums in game-like interactions, and to add more power to the protection and inheritance of intangible cultural heritage.

\end{document}